\documentclass[opre,nonblindrev]{informs3} 

\DoubleSpacedXI 

\usepackage{endnotes}
\let\footnote=\endnote

\usepackage{natbib}
 \bibpunct[, ]{(}{)}{,}{a}{}{,}%
 \def\bibfont{\small}%
 \def\newblock{\ }%
\TheoremsNumberedThrough     
\ECRepeatTheorems

\EquationsNumberedThrough    

\MANUSCRIPTNO{} 

\usepackage{xspace}
\newcommand{\TEOQ}{EOQ\xspace} 
\newcommand{\EOQ}[1]{#1^\mathrm{eoq}} 
\newcommand{\CEOQ}[3]{\mathrm{EOQ}_{#2,#3}(#1)} 

\newcommand{\SH}[3]{\mathrm{DC}(#1,#2;#3)} 
\newcommand{\JC}[3]{\mathrm{JC}_{#3}(#1,#2)} 
\newcommand{\M}{M}

\newcommand{\JRP}{JRP\xspace}
\newcommand{\JRPF}[2]{\mathrm{JRP}(#1,#2)} 
\newcommand{\JRPGIF}[2]{\mathrm{JRP}(#1,#2)} 
\newcommand{\JRPMLU}{\mathrm{JRP}_{M,U,L}\xspace} %

\newcommand{\JRPMR}{\mathrm{JRP}^{\mathrm{red}}\xspace}
\newcommand{\JRPGICF}{{aperiodic \JRP}\xspace}
\newcommand{\JRPGI}{periodic \JRP}

\newcommand{\NPFact}{\textsc{RangeDivisor}\xspace}
\newcommand{\PROBLEM}{\Pi}

\newcommand{\NP}  {{NP}\xspace}

\newcommand{\lb}  {l\xspace} 
\newcommand{\ub}  {h\xspace} 

\newcommand{\floor}[1]  {\left\lfloor#1 \right \rfloor } 
\newcommand{\ceil}[1]   {\left\lceil #1 \right \rceil }  

\newcommand{\lcm}{\mathbf{lcm}}
\renewcommand{\gcd}{\mathbf{gcd}}
\newcommand{\ZZ}{\mathbb{Z}}
\newcommand{\RR}{\mathbb{R}}
\newcommand{\QQ}{\mathbb{Q}}
\newcommand{\NN}{\mathbb{Z}_+}

{\theoremstyle{THkey}\newtheorem{mytheorem}{XXXXX}}

\begin{document}


\RUNAUTHOR{A.S.\ Schulz and C.\ Telha}

\RUNTITLE{Integer factorization and Riemann's hypothesis: Why two-item joint replenishment is hard}

\TITLE{Integer factorization and Riemann's hypothesis:\\ Why two-item joint replenishment is hard}

\ARTICLEAUTHORS{%
\AUTHOR{Andreas S.\ Schulz}
\AFF{School of Management and Department of Mathematics, Technische Universit\"at M\"unchen, M\"unchen, Germany\\ \EMAIL{andreas.s.schulz@tum.de}} 
\AUTHOR{Claudio Telha}
\AFF{Facultad de Ingenier\'ia y Ciencias Aplicadas, Universidad de los Andes, Santiago de Chile, Chile\\ \EMAIL{catelha@miuandes.cl}}
} 

\ABSTRACT{%
Distribution networks with periodically repeating events often hold great promise to exploit economies of scale.  
Joint replenishment problems are a fundamental model in inventory management, manufacturing, and logistics that capture these effects.  However, finding an efficient algorithm that optimally solves these models, or showing that none may exist, has long been open, regardless of whether empty joint orders are possible or not.  In either case, we show that finding optimal solutions to joint replenishment instances with just two products is at least as difficult as integer factorization.  To the best of the authors' knowledge, this is the first time that integer factorization is used to explain the computational hardness of any kind of optimization problem.  Under the assumption that Riemann's Hypothesis is correct, we can actually prove that the two-item joint replenishment problem with possibly empty joint ordering points is NP-complete under randomized reductions, which implies that not even quantum computers may be able to solve it efficiently.  By relating the computational complexity of joint replenishment to cryptography, prime decomposition, and other aspects of prime numbers, a similar approach may help to establish (integer factorization) hardness of additional open periodic problems in supply chain management and beyond, whose solution has eluded standard methods.
}%



\maketitle

%


\section{Introduction}
In industrial activities, the simplicity and conciseness of periodic (or cyclic) schedules make them ubiquitous and prevalent
(e.g., \citet{Hanen_Munier_Curie_1994, Crama_Kats_2000, Mendez_Cerda_Grossmann_Harjunkoski_Fahl_2006, Dawande_Geismar_Pinedo_Sriskandarajah_2009, Levner_Kats_Alcaide_2010}).  A simple, periodic schedule will usually overtake any complex schedule as it eases management, understanding, administration, and accounting, even if it is sub-optimal.  However, one may want to align different multiple periodic processes in time, to possibly benefit from economies of scale.  A concrete example is the 
replenishment of inventory~\citep{Muckstadt_Roundy_1993}, where the simultaneous replenishment of more than one product is an opportunity to share transportation or handling costs.  Regardless of the concrete application in mind, one might ask: How difficult is it to optimally coordinate a set of periodic activities?  In other words, what is the computational effort required to determine optimal periods of a collection of activities that are driven not only by individual incentives, but also by potential collective gains, which would require some form of synchronization.  In this paper, we establish the computational hardness of coordinating {\em two} periodic activities, which is arguably the simplest form of coordination that one may encounter.  More specifically, we prove that coordinating two (and, hence, any number of) activities is at least as hard as integer factorization, when the periods are integers (i.e., we have discrete period processes).  
Joint replenishment is serving as the representative problem for our approach.  Before we describe the model, we quickly recall the basics for the single-item setting.

\subsection{Economic Order Quantity}
\begin{figure}
\FIGURE
{\includegraphics[width=12cm]{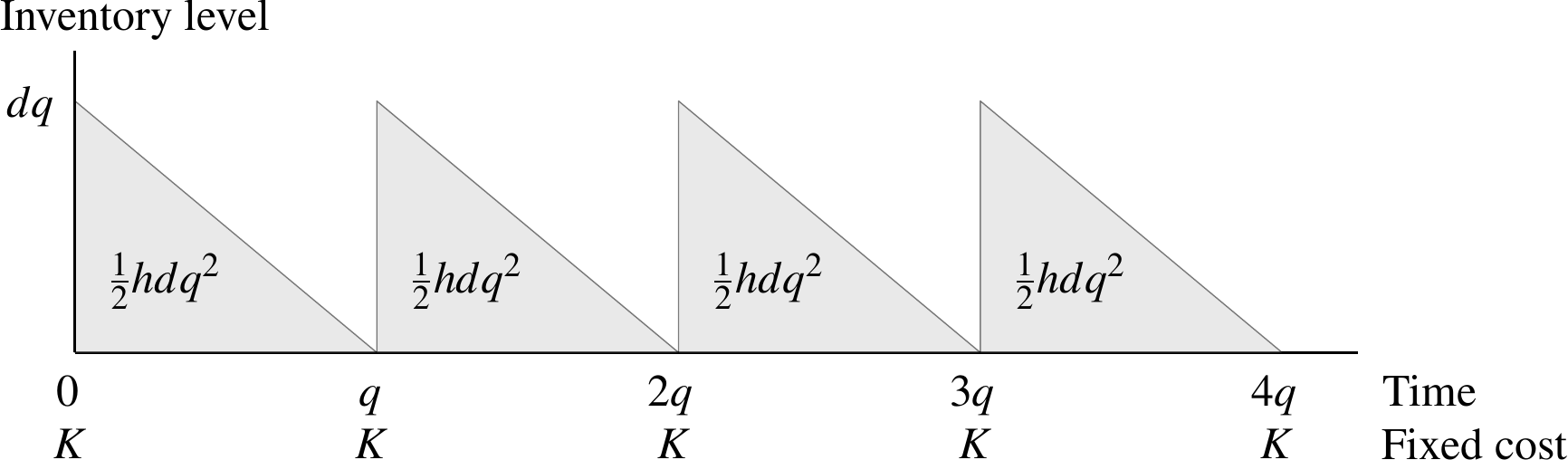}}
{Economic Order Quantity Model \label{fig:eoq}}
{}
\end{figure}

The Economic Order Quantity (\TEOQ) is one of the earliest, fundamental results in inventory theory~\citep{Harris_1913}.  It specifies how much should be ordered so as to minimize the average ordering and holding cost of a single item over an infinite time horizon, under a number of simplifying assumptions.  Specifically, inventory is stocked to satisfy a constant demand $d$ per unit of time; a fixed cost $K$ is charged at each reordering point; a holding cost $h$ is incurred for each unit of item stored per unit of time. 
The optimal solution is periodic and has a reordering time $q = \sqrt{K/H}$, which minimizes $\CEOQ{q}{K}{H}\equiv K/q + Hq$, where  $H = h d / 2$ (see Figure~\ref{fig:eoq}). 

\subsection{Joint Replenishment Problems}
A frequently occurring multi-product extension of this problem is to determine a min-cost replenishment plan for several products ordered from the same supplier.  Ordering different products at the same point in time often results in substantial savings, compared with ordering the same products independently.  This, so-called, joint replenishment problem (\JRP) has been the subject of intensive research since the early sixties.
The incentive to coordinate is driven by an {\em additional}, {\em item-independent} cost associated with every possible ordering point.  It can arise in two different forms: 1) costs which are incurred periodically (this is the most studied variant, see the surveys by \cite{Goyal_Satir_1989, Khouja_Goyal_2008,Bastos_Mendes_Nunes_Melo_Carneiro_2017}); and 2) costs that are incurred only whenever one or more products are actually ordered (e.g.,  \cite{Dagpunar_1982,Porras_Dekker_2008,
Cohen-Hillel_Yedidsion_2018b}).  For instance, suppose orders may be placed every other day, and there are two products with replenishment periods $q_1=4$ and $q_2=6$, respectively.  In the first cost model, the item-independent joint ordering cost would be charged at times $0, 2, 4, 6, 8, 10, 12$, and so on.  In the second cost model, the joint ordering cost would only be imposed at actual order times, i.e., at times $0, 4, 6, 8, 12$, etc., skipping the empty orders at times $2$, $10$, and so forth.   We refer to the \JRP with cost model 1) as {\em periodic} \JRP, and to the \JRP with cost model 2) as {\em aperiodic} \JRP. (The periodic \JRP is also known as the {\em general integer model}, and the \JRPGICF\ is known as the {\em general integer model with correction factor}.)  These two accounting models are substantially different.
A nice comparison of the two variants was given by~\cite{Porras_Dekker_2008}.

\subsection{Main Results}
Regardless of how extra ordering costs are accounted for, no polynomial-time algorithm for computing an optimal replenishment plan is known, let alone closed-form solutions, unlike in the \TEOQ case.  Scores of research~\citep[e.g.,][]{Silver_1976, Jackson_Maxwell_Muckstadt_1988, Kaspi_Rosenblatt_1991, Lu_Posner_1994, Klein_Ventura_1995, Wildeman_Frenk_Dekker_1997, Teo_Bertsimas_2001, Nilsson_Segerstedt_van_der_Sluis_2007, Praharsi_Purnomo_Wee_2010} 
have been devoted to developing heuristics (which in general fail to find optimal solutions) and enumerative algorithms (which are computationally prohibitive for large instances)---approaches that are usually reserved for problems that are provably hard.  However, until very recently there was no theoretical evidence that finding efficient exact algorithms might be impossible.

In an earlier stage of this paper, we showed that the aperiodic \JRP is, in the sense of Turing reductions, as least as hard as integer factorization \citep{Schulz_Telha_2011}.  Given a composite number, Integer Factorization is the problem of decomposing that number into the product of two smaller numbers.  Even though integer factorization is unlikely to be NP-hard (for instance, it belongs to NP {\em and} co-NP), it is widely believed to be a ``computationally hard'' problem in practice (e.g., \cite{Lenstra_2000}).  In fact, this belief fuels the assumption that RSA \citep{Rivest_Shamir_Adleman_1978} and other cryptographic systems (and, hence, online banking and other online transactions) are secure.

Here, we introduce the notion of {\em integer factorization hardness} more formally.  We are convinced that it will have important bearing on our understanding of several other optimization problems.  Indeed, we will prove integer factorization hardness for the periodic \JRP, too.  Its computational complexity had been open for a long time. Moreover, we will also use this problem to demonstrate that our approach can sometimes lead to ordinary NP-hardness results, which arise from a computationally harder version of integer factorization.


Here is an overview of our integer factorization hardness results for the \JRP:

\begin{theorem}\label{thm:gicf:hardnesss}${}$
\begin{itemize}
\item[{\rm (a)}] Suppose there is a polynomial-time algorithm for the {\em aperiodic} \JRP with two items.  Then there is a polynomial-time algorithm for integer factorization.
\item[{\rm (b)}] Suppose there is a polynomial-time algorithm for the {\em periodic} \JRP with two items.  Then there is a polynomial-time algorithm for integer factorization.
\end{itemize}
\end{theorem}

As mentioned before, the computational complexity of both variants of the \JRP had been open.  After we announced our first result (i.e., Theorem~\ref{thm:gicf:hardnesss}(a)) in \cite{Schulz_Telha_2011}, \citet{Cohen-Hillel_Yedidsion_2018b} proved that the aperiodic \JRP with an {\em arbitrary} number of items is strongly NP-hard.  (They actually refer to this problem as the {\em periodic} \JRP in their paper; even though it may cause some confusion at first, we prefer to stick with our versions of {\em periodic} vs.\ {\em aperiodic}, to highlight whether the additional item-independent ordering cost arises in periodic intervals or not.)  Their proof is remarkable, and quite involved; it uses nontrivial facts from number theory, is 16 pages long, and does rely on blowing up both the cost coefficients and the number of items.  In contrast, our construction is short, elementary, works for the two-item case, and is not specific to the aperiodic \JRP.  Furthermore, we not only give a succinct proof that the periodic \JRP is integer factorization hard as well, even if there are just two items (i.e., Theorem~\ref{thm:gicf:hardnesss}(b)), but also push it further:

\begin{theorem}\label{thm:gi:hardness}
Assume one can prove the Riemann hypothesis.  Then, the periodic \JRP with {\em two} items is NP-hard, under randomized reductions.
\end{theorem}

The Riemann hypothesis, formulated by Riemann (1859), states that the real part of all nontrivial zeros of the Riemann zeta function is equal to $1/2$.  It is contained in Hilbert's (1900) original list of 23 unsolved mathematical problems, and it is one of the seven Millennium Problems of the Clay Mathematics Institute \citep{Jaffe_2006}.  Whether true or not, either way a resolution of the Riemann hypothesis would have important implications on the distribution of prime numbers (see, e.g., \citet{Bombieri_2000}).


In Sections~\ref{sec:gicf} and \ref{sec:gi} we explain all arguments required to prove Theorems~\ref{thm:gicf:hardnesss} and \ref{thm:gi:hardness}, respectively.
Section~\ref{sec:conclusion} contains our concluding remarks.\\

\section{Two-Item \JRP Is At Least As Hard As Factoring} \label{sec:gicf}
Integer factorization is one of the oldest problems in mathematics.  Still, no classical (i.e., non-quantum) polynomial-time algorithm is known, even though such an algorithm does exist for testing whether a given number is prime~\citep{Agrawal_Kayal_Saxena_2004}.  However, \cite{Shor_1999} discovered a polynomial-time quantum algorithm to solve integer factorization.  In practice, though, integer factorization remains a difficult computational problem~\citep{Lenstra_2000}. For instance, the RSA cryptographic protocol~\citep{Rivest_Shamir_Adleman_1978} keeps data secure by protecting it behind a public and a private key.  Since the private key is a number immediately accessible to anyone capable of factoring large coprimes efficiently, the difficulty of integer factorization is what keeps the data protected.  (Recall that two integers are coprime(s) if the only positive integer dividing both of them without any remainder is~$1$.)  How is integer factorization related to coordination problems?  Our proofs will rely on a standard concept: a (Turing) \emph{reduction} from a problem $A$ (e.g., integer factorization) to a problem $B$ (e.g., a coordination problem) is an algorithm that solves any one instance of $A$ using one or more oracle calls to a hypothetical algorithm that can solve instances of $B$.  When the reduction itself is polynomial, we say that $B$ is at least as difficult to solve as $A$.  More formally, we have the following definition.

\begin{definition}
A computational problem $\Pi$ is {\em integer factorization hard} if there is a polynomial-time Turing reduction from Integer Factorization to $\Pi$.
\end{definition}

In our reductions to the different versions of \JRP, the least common multiple ($\lcm(\cdot)$) of two periods $q_1$ and $q_2$ can be seen as a measure of sync: it can range from perfect synchronization ($q_1=q_2$) to not being in sync at all ($q_1$ is coprime with $q_2$).  At the same time, computing the least common multiple relates to integer factorization: $\lcm(q_1,q_2)$ gives access to a divisor of both $q_1$ and $q_2$. (Recall that the product of the least common multiple and the greatest common divisor of two integers is always equal to the product of the two integers.)

\subsection{The Reduction For The Aperiodic \JRP}
\begin{figure}
\FIGURE
{\includegraphics[width=12cm]{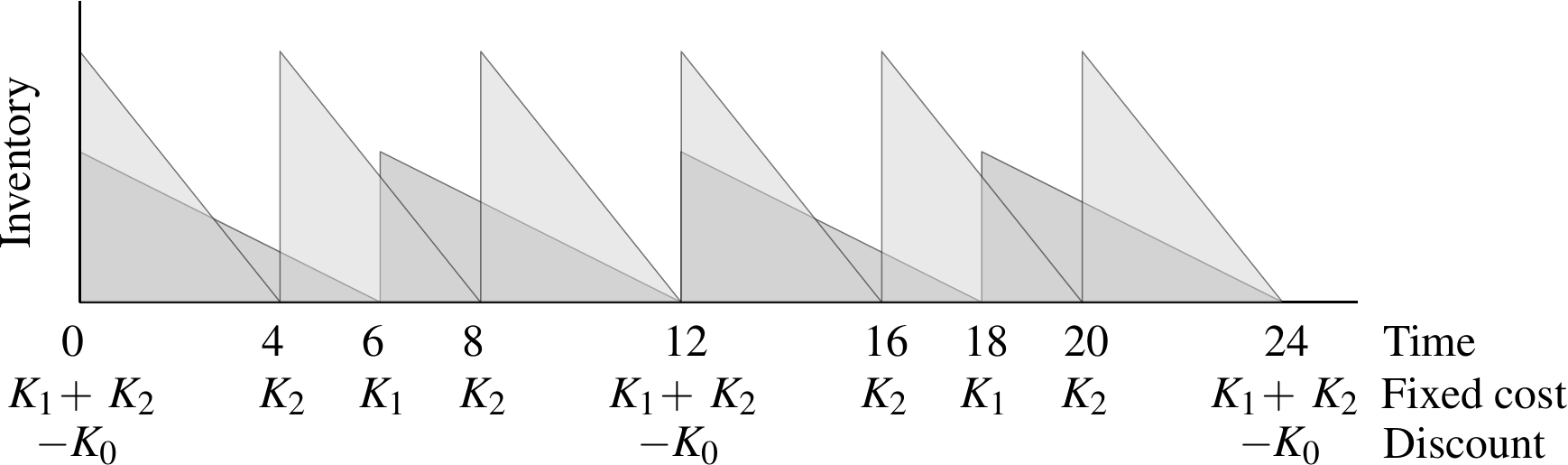}}
{Replenishment plan $q_1=6$, $q_2=4$ for a $2$-product instance of the \JRPGICF \label{fig:jrp:2prod}}
{}
\end{figure}

We work with an equivalent reformulation of the \JRPGICF, which applies only to two-item instances (see Figure~\ref{fig:jrp:2prod}). In this reformulation, the periodic ordering of two products 1 and 2 is still subject to individual fixed costs $K_1, K_2$ and holding costs $H_1, H_2$ according to the \TEOQ model, but the fixed costs are subject to a discount $K_0 \leq K_1, K_2$ each time both products are jointly replenished.  Put differently, $K_1-K_0$ and $K_2-K_0$ correspond to the individual fixed costs in the original formulation.  Given the lengths of the replenishment cycles $q_1\in \ZZ^+$ and $q_2\in \ZZ^+$, the discount can be expressed as $\SH{q_1}{q_2}{K_0} \equiv K_0/\lcm(q_1,q_2)$.
The problem becomes to determine $q_1$ and $q_2$ so as to minimize
\begin{equation}\label{eq:jrp:gicf:disc}
\JRPF{q_1}{q_2} \equiv \CEOQ{q_1}{K_1}{H_1} + \CEOQ{q_2}{K_2}{H_2} - \SH{q_1}{q_2}{K_0}.
\end{equation}

\subsubsection*{From Coordination to Integer Factorization.}
It is well known that finding the integer factorization of a composite number $M$ reduces to the problem of finding one of its  non-trivial divisors~\citep[Chapter~10.6, pp. 222]{Arora_Barak_2009}. We now outline a method to find one such divisor using the solution to a \JRPGICF instance with only two items.

\begin{figure}
\FIGURE
{\includegraphics[scale=0.8]{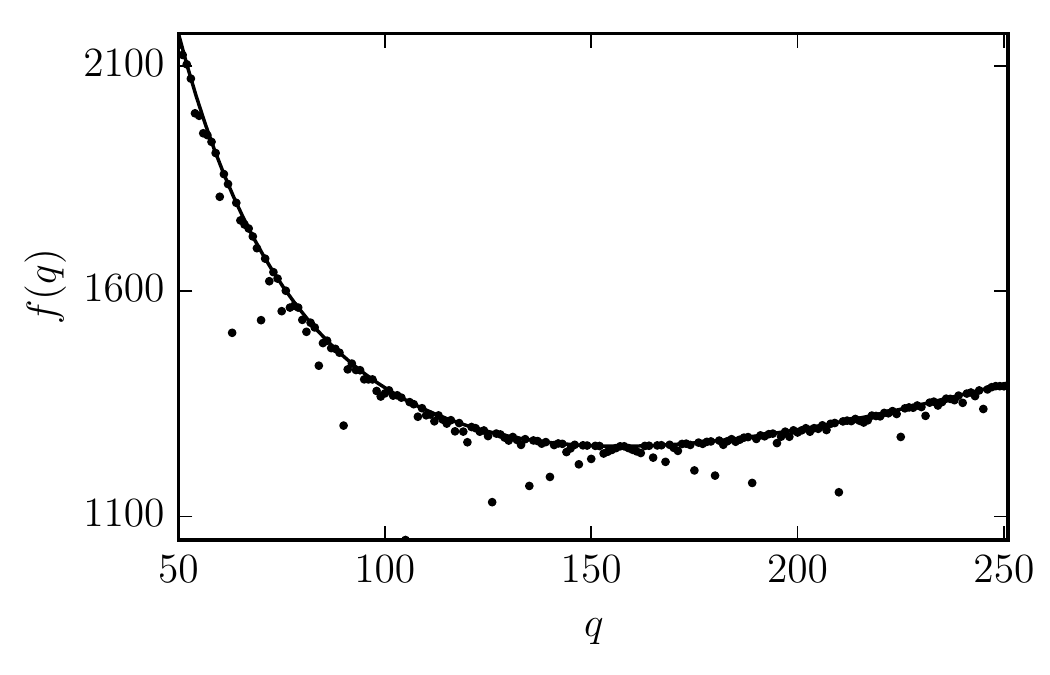}}
{\label{fig:gicf:func}}
{An example of the function $f(q) = \CEOQ{q}{K_1}{H_1} - \SH{q}{M}{K_0}$ for $M=315$, with parameters $K_1=K_0=M(M-1)$ and $H_1=4$. The function takes values below the convex curve at multiples of 5, 7 or 9.}
\end{figure}

The non-coordinated \TEOQ cycle lengths $(\EOQ{q_1},\EOQ{q_2})$ are, in general, sub-optimal solutions to \eqref{eq:jrp:gicf:disc}.  One approach to find a possibly better solution is to modify $(\EOQ{q_1},\EOQ{q_2})$ so that the increase in discounts overcompensates the total increase in individual costs.
Suppose 
that we alter $q_1$ while keeping $q_2$ fixed at $ \EOQ{q_2} = M$. An example of how the function $\CEOQ{q_1}{K_1}{H_1} - \SH{q_1}{M}{K_0}$ looks like is given in Figure~\ref{fig:gicf:func}. When $q_1$ is coprime with $M$, the function takes values aligned with those of a convex ``quadratic'' function. When $q_1$ is non-coprime with $M$, the function jumps down due to the higher discount.

Our goal is to set up a two-item instance of the  \JRPGICF whose optimal solutions satisfy $q_2=\EOQ{q_2}=\M$ while achieving some degree of synchronization, meaning that $1< \gcd(q_1,M) < M$. The greatest common divisor $\gcd(q_1,M)$, which can be computed in polynomial time, is then a non-trivial divisor of $\M$, as needed.

\subsubsection*{Detailed Reduction.}
Given a composite number $M$, let us study the minimization of \eqref{eq:jrp:gicf:disc} for the instance $\JRPF{q_1}{q_2}$ defined by $K_1 = K_0 = M(M-1)$, $H_1=4$, $K_2=M^3(M^2-1)$, and $H_2=M(M^2-1)$.
Because the period $q_2$ is discrete and the weight of item 2 dominates the overall costs, choosing anything but $q_2=\EOQ{q_2}=M$ will lead to a sub-optimal solution.
With $q_2$ fixed to $q_2=M$, $q_1$ must minimize
\begin{equation}
\begin{tabular}{r l}\label{eq:gicf:redfunc}
$\min\phantom{dd}$ &   $M(M-1) \left(\frac{1}{q_1} - \frac{1}{\lcm(q_1,M)} \right)+4q_1$\\[0.2cm]
s.t.\phantom{\phantom{dd} } & $q_1\in\ZZ^+$.
 \end{tabular}
\end{equation}
An example of the function to be minimized is given in Figure~\ref{fig:gicf:func}. We already used this figure to argue that $1 < \gcd(q_1,M)$, and one can also use it to argue that $\gcd(q_1,M)< M$. Indeed, the minimizer of the convex curve will always be attained at $\frac{M-1}{2}$, and this is sufficiently far from $M$ that not the highest of the discounts could make $q_1=M$ the optimum of~\eqref{eq:gicf:redfunc}.  The formal lemma is the following:
\begin{lemma}\label{thm:gicf:reduction}
For any odd composite number $M$, there is a two-item instance $\min\{\JRPF{q_1}{q_2} : q_1,q_2\in\ZZ^+ \}$ of the \JRPGICF whose optimal solutions $(q_1,q_2)$ satisfy $q_2=M$ and $1< \gcd(q_1,M)<M$. In addition, the size of the coefficients of this instance is at most $5\log(M)$.
\end{lemma}

\proof{Proof.}
The instance $\JRPF{q_1}{q_2} = \CEOQ{q_1}{K_1}{H_1} + \CEOQ{q_2}{K_2}{H_2} - \SH{q_1}{q_2}{K_0}$
with parameters $K_1 = K_0 = M(M-1)$, $H_1=4$, $K_2=M^3(M^2-1)$, and $H_2=M(M^2-1)$ satisfies the size requirements stated in the lemma. We split the rest of the proof into two parts.

\paragraph{Optimal replenishment cycle for item 2.}
We first show that any minimizer of $\JRPF{q_1}{q_2}$ satisfies $q_2=M$: The function $\CEOQ{q_2}{K_2}{H_2}$ is minimized at $q_2=\EOQ{q_2}=M$. By convexity and integrality,
 \begin{displaymath}
\JRPF{q_1}{q_2} \geq 4q_1+ \min \{\CEOQ{M-1}{K_2}{H_2}, \CEOQ{M+1}{K_2}{H_2} \}
 \end{displaymath}
for any $q_2 \neq M$. Since
\begin{displaymath}
\CEOQ{M\pm 1}{K_2}{H_2} = \CEOQ{M}{K_2}{H_2}  + \frac{H_2}{M\pm 1} \geq  \CEOQ{M}{K_2}{H_2} + K_0
\end{displaymath}
we get that $\JRPF{q_1}{q_2} \geq \CEOQ{M}{K_2}{H_2} + K_0 + 4$. This lower bound is already greater than the value of $\JRPF{1}{M}$, implying that any $q_2\neq M $ is strictly sub-optimal.

\paragraph{Optimal replenishment cycle for item 1.}
After substituting $q_2=M$, the \JRP instance reduces to the optimization problem \eqref{eq:gicf:redfunc}
with objective $\JRPMR(q_1) \equiv  M(M-1) \left(\frac{1}{q_1} - \frac{1}{\lcm(q_1,M)} \right)+4q_1$ over the single variable $q_1$.
Our goal in the second part of the proof is to show that for any odd composite number $M$, the minimizers $q_1^*$ 
satisfy $1 < \gcd(q_1^*,M)< M$.

First, let us show that $\gcd(q_1^*,M) > 1$. For any $q_1'$ coprime with $M$, we can bound $$\JRPMR(q_1')\geq \min_{q_1\in
\RR}\left\{\frac{(M-1)^2}{q_1}+ 4q_1 \right\} =4(M-1).$$
Consequently, we just need to show that  $\JRPMR(q_1^*) < 4(M-1)$. Let $p$ be a non-trivial divisor of $M$ such that $3 \leq p < M$, and let $q$ be an arbitrary multiple of $p$. The discount amount satisfies $\SH{q}{M}{K_0} = M(M-1)/\lcm(q,M) \geq p(M-1)/q$ and, therefore,
$$\JRPMR(q)  \leq \frac{(M-1)(M-p)}{q}+4q .$$

This upper bound, as a continuous function of $q$, is strictly convex. It attains the value $4(M-1)$ at the two solutions of $2q = (M-1) \pm \sqrt{(M-1)(p-1)}$, and is strictly smaller than $4(M-1)$ in between. In particular, the upper bound is smaller than $4(M-1)$ in the interval
$$\left[\frac{M-1}{2} - \frac{p-1}{2}, \frac{M-1}{2}
+\frac{p-1}{2}\right] \cap \ZZ^+,$$ which contains exactly $p$ integers. One of them must be a multiple $q$ of $p$, therefore it must satisfy $\JRPMR(q)<4(M-1)$.

Next, let us show that $\gcd(q_1^*,M)< M$. Indeed, if $\gcd(q_1,M) = M$, then $q_1$ is a multiple of $M$ and there is a discount at each replenishment of item~1. It easily follows that $\JRPMR(q_1)= 4q_1 \geq 4M$ for any such $q_1$. This cannot be an optimal solution since $\JRPMR((M-1)/2) \leq 4(M-1)$.
\Halmos
\endproof

\smallskip

In summary, we have reduced integer factorization to the \JRPGICF: to find a non-trivial divisor of an odd composite number $M$ we solve the two-item instance, and then we return the greatest common divisor of its optimizers $q_1$ and $q_2$. Even numbers, which are not covered by Lemma~\ref{thm:gicf:reduction}, are easily recognizable and have an immediate non-trivial divisor. To obtain the full factorization of $M$, once a non-trivial divisor $d$ of  $M$ is found, we recursively find non-trivial divisors of $d$ and $M/d$ as long as they are composite numbers.  Recall that testing whether a number is prime can be solved in polynomial time \citep{Agrawal_Kayal_Saxena_2004}. This concludes the proof of Theorem~\ref{thm:gicf:hardnesss}(a).

\begin{repeattheorem}\label{thm:gicf:hardnesss}
{\rm (a)} Suppose there is a polynomial-time algorithm for the aperiodic \JRP with two items.  Then there is a polynomial-time algorithm for integer factorization.
\end{repeattheorem}

\subsection{The Reduction For The Periodic \JRP}
We now turn to the \JRPGI, and we again focus on the simplest possible instances of relevance, i.e., the two-item case.  The joint ordering cost $K_0$ is now incurred every $\rho$ units of time, for some integer $\rho \ge 1$. (Formally, to be consistent with the existing literature, $\rho$ needs to be a multiple of some given parameter $B$.  In this paper, we assume throughout that $B = 1$.)   Each item $i=1,2$ follows an \TEOQ model with parameters $(K_i,H_i)$, and its ordering period is a multiple of $\rho$. Assuming that $K_0>0$, one has $\rho=\gcd(q_1,q_2)$ for every optimal solution, so the joint ordering cost can be expressed as $\JC{q_1}{q_2}{K_0} \equiv K_0/\gcd(q_1,q_2)$. 
The objective becomes to minimize:
\begin{equation}\label{eq:gi:def}
\JRPGIF{q_1}{q_2} \equiv \CEOQ{q_1}{K_1}{H_1} + \CEOQ{q_2}{K_2}{H_2} + \JC{q_1}{q_2}{K_0}.
\end{equation}

The difference between this model and the \JRPGICF is how joint ordering costs are accounted for.  This difference renders the two problems substantially different from one another.  No hardness results were previously known for the {\JRPGI}.

Given a composite number $M$, let us study the minimization of \eqref{eq:gi:def} for the instance $\JRPGIF{q_1}{q_2}$ defined by $K_0=M$, $K_1=0$, $H_1=1$, $H_2=M(M+1)$, and $K_2= M^3(M+1)$.
Similar to the reduction for the aperiodic \JRP, the weight of item 2 dominates the overall costs, so that choosing anything but $q_2 =\EOQ{q_2} =M$ will lead to sub-optimal solutions. Taking this ``variable fixing'' into account, the problem reduces to
\begin{equation}
\begin{tabular}{r l}
$\min$ \phantom{dd} &$\frac{M}{\gcd(q_1,M)}+ q_1$ \label{eq:gicfIF:redfunc} \\
s.t. \phantom{dd} & $q_1 \in \ZZ^+$.
\end{tabular}
\end{equation}

In this case, the greatest common divisor $\gcd(q_1 ,M)$ appears directly in the objective function.  More formally, we have the following lemma.  Its proof follows the logic of the proof of Lemma~\ref{thm:gicf:reduction}, except that it turns out to be even simpler.

\begin{lemma}\label{thm:gi:if:reduction}
For any odd composite number $M$, there is a two-item instance $\min\{\JRPGIF{q_1}{q_2}: q_1,q_2\in\ZZ^+ \}$ of the periodic \JRP whose optimal solutions $(q_1,q_2)$ satisfy $q_2=M$ and $1< \gcd(q_1,M)<M$. In addition, the size of the coefficients of this instance is at most $5\log(M)$.
\end{lemma}

\proof{Proof.}
Given $M$, the instance defined by $K_0=M$, $K_1=0$, $H_1=1$, $H_2=M(M+1)$, and $K_2= M^3(M+1)$  satisfies the stated size requirements. 

\paragraph{Optimal replenishment cycle for item~2.}
We first show that any minimizer of $\JRPGIF{q_1}{q_2}$ satisfies $q_2=M$. Similar to the proof of Lemma~\ref{thm:gicf:reduction}, we have that for any $q_2\neq M$:
\begin{displaymath}
\CEOQ{q_2}{K_2}{H_2} \geq \CEOQ{M\pm 1}{K_2}{H_2} = \CEOQ{M}{K_2}{H_2}  + \frac{H_2}{M\pm 1} \geq  \CEOQ{M}{K_2}{H_2} + M.
\end{displaymath}
Hence, $\JRPGIF{q_1}{q_2} > M + \CEOQ{M}{K_2}{H_2} + 1 $ for any $q_2\neq M$. This lower bound is already greater than the value of $\JRPGIF{M}{M}=M+ \CEOQ{M}{K_2}{H_2} + 1$, implying that any $q_2\neq M $ is strictly sub-optimal.

\paragraph{Optimal replenishment cycle for item~1.}
After substituting $q_2=M$, the periodic \JRP instance reduces to the optimization problem \eqref{eq:gicfIF:redfunc} on the single variable $q_1$.
%
%
Because $M$ is an odd composite number, it has a non-trivial divisor $p$ with $3 \leq p < M/3$. For such $p$, we have, for the objective function value of \eqref{eq:gicfIF:redfunc}, that $\frac{M}{\gcd(p,M)}+ p \leq 2M/3$.

On the other hand, when $\gcd(q,M)=1$ or $\gcd(q,M)=M$, it is immediate that the objective function value of \eqref{eq:gicfIF:redfunc} is at least $M$.  It follows that the optimal solutions $q_1$ of \eqref{eq:gicfIF:redfunc} satisfy $1 < \gcd(q_1,M)< M$, as claimed.
\Halmos
\endproof

Obviously, Lemma~\ref{thm:gi:if:reduction} can be used to create the desired reduction from Integer Factorization to the \JRPGI.  Note that the choice of $K_1=0$ simplifies the expressions above, but it is not necessary (e.g., $K_1=1$ also works).  
We have completed the proof of Theorem~\ref{thm:gicf:hardnesss}(b).

\begin{mytheorem}[Theorem~1]
{\rm (b)} Suppose there is a polynomial-time algorithm for the periodic \JRP with two items.  Then there is a polynomial-time algorithm for integer factorization.\\
\end{mytheorem}

%

\section{Going Beyond Integer Factorization Hardness}\label{sec:gi}
For the \JRPGI, the proof technique introduced in Section~\ref{sec:gicf} can be enhanced to establish an even stronger result.  For this, we need to formally express integer factorization as a decision problem (e.g., \citet[Chapter~2.1, pp. 40]{Arora_Barak_2009}): given integers $M \geq U \geq L > 0$, decide if $M$ has a prime factor $p$ in the interval $[L,U]$. A seemingly technical detail in this definition is crucial to the following discussion: by {\em not} requiring $p$ to be a prime number, we obtain a problem that seems to be much harder than integer factorization. Consider the following decision problem:
\smallskip
\begin{quote}
\NPFact: Given integers $M \geq U \geq L > 0$ satisfying $L+U< 2\sqrt{2LU}$, decide if $M$ has a divisor in the interval $[L,U]$.
\end{quote}
\smallskip
\noindent A result by \cite{Shor_2011} implies that, for every decision problem $\PROBLEM$ in the class \NP, there is a polynomial-time computable map $f:\textrm{inputs}(\PROBLEM) \to \textrm{inputs}(\NPFact)$, so that, with high probability, $x\in \PROBLEM \Rightarrow f(x) \in \NPFact$, and $x\notin \PROBLEM \Rightarrow f(x) \notin \NPFact$.  Minding the failure probability, this is a reduction from $\PROBLEM$ to \NPFact, and it shows that \NPFact is complete in the class \NP, under randomized reductions. Although \NP-completeness under randomized reductions is not frequently used, it is arguably as good as ordinary \NP-completeness as a complexity measure~\citep[Chapter~7.6, pp. 138]{Arora_Barak_2009}.  We provide a self-contained proof of Shor's result in the appendix because the implication is, in fact, not direct.  This is the part where Riemann's hypothesis is needed.

By reducing \NPFact to the \JRPGI with two items, we show that the two-item \JRPGI is also complete for the class \NP, under randomized reductions.  For this purpose, let us formally state the two-item \JRPGI as a decision problem: Given $K_0, K_1, K_2$, $H_1, H_2$, as well as a threshold value $z \in \QQ$, is there a solution $q_1, q_2$ of cost at most $z$?

\subsection{The Reduction}
Consider an instance of $\NPFact$, i.e., let $M \geq U \geq L > 0$ be three integers satisfying $2\sqrt{2LU}> L+U$.  We want to build a two-item instance $\JRPMLU$ of the \JRPGI and a threshold $z_{M,L,U}$ so that there is a divisor of $M$ in the interval $[L,U]$ if and only if the joint replenishment problem has an optimum with cost less than or equal to the threshold $z_{M,L,U}$.

Some of the ideas from the integer factorization reductions come now into play.  The instance $\JRPMLU$ will assign a large weight to item 2,  so that $q_2=M$ is enforced in every optimal solution. We also set $K_1=0$, $H_1=1$, as well as $K_0=LU$.
An optimal period for item 1 must therefore solve:
\begin{equation}
\begin{tabular}{r l}
$\min$ \phantom{dd} &$\frac{LU}{\gcd(q_1,M)} + q_1$ \label{eq:gi:redfunc} \\[0.2cm]
s.t. \phantom{dd} & $q_1 \in \ZZ^+$.
\end{tabular}
\end{equation}

\begin{figure}
\FIGURE
{\includegraphics[scale=0.8]{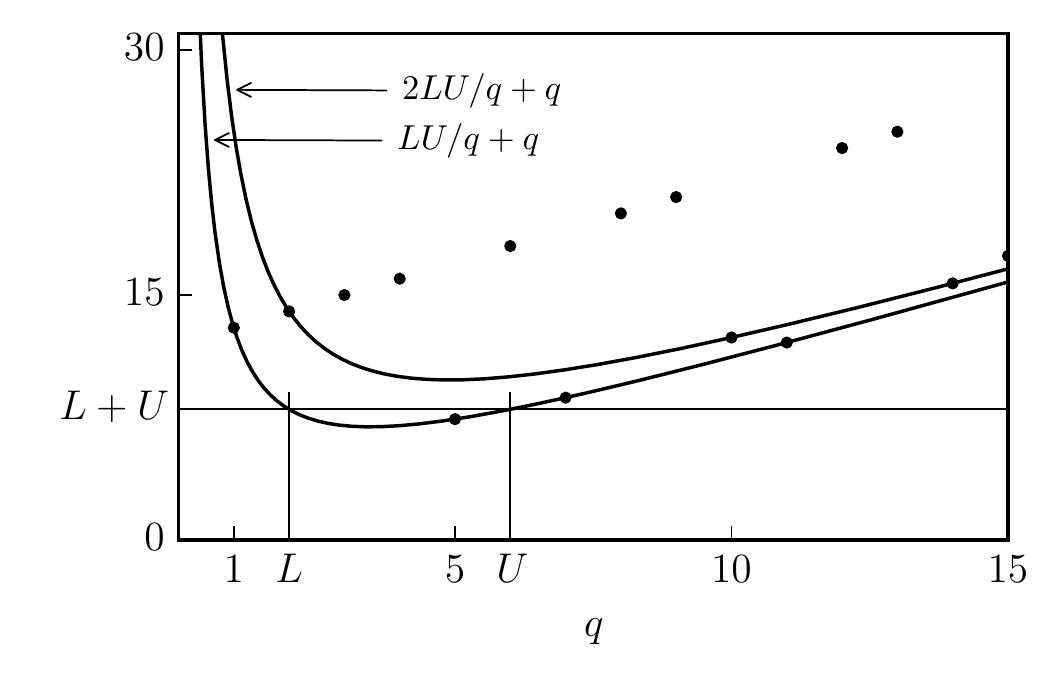}}
{}
{The discrete points correspond to the function $\frac{LU}{\gcd(q,M)} + q$, when $M=385$, $L=2$ and $U=6$.\label{fig:gi:func}}
\end{figure}

An example of the kind of function to be minimized is given in  Figure~\ref{fig:gi:func}. Let us focus on the range $\{1,\ldots,M\}$, in which an optimum must lie. The function $\ub(q)\equiv LU/q + q$ equals the objective  when $q_1$ is a divisor of $M$. The function $\lb(q)\equiv 2LU/q + q$ bounds the objective from below when $q_1$ is not a divisor of $M$. Intuitively, the gap between the convex functions $\lb$ and $\ub$ is sufficiently big to separate the optimal value $z^*$ of Problem \eqref{eq:gi:redfunc} depending on whether the interval $[L,U]$ contains a divisor of $M$: if yes, then $z^* \leq  L+U$. If not, then $z^*> L+U$.  The details are in the proof of the following lemma.

\begin{lemma}\label{lemmaJRPGI}
For any instance $M \geq U \geq L > 0$ of $\NPFact$, one can construct, in polynomial time, a two-item
instance  $\min\{\JRPMLU(q_1,q_2): q_1,q_2\in\NN \}$ of the \JRPGI with the following properties: there is a divisor $q$ of $M$ in the interval $[L,U]$ if and only if  the optimal value of $\JRPMLU \leq L+U$. In addition, the size of the coefficients of the two-item instance is polynomial in $\log(M)$.
\end{lemma}

\proof{Proof.}
Let $M, L, U$ be an instance of \NPFact.  We construct a two-item instance of the \JRPGI as follows.
We fix $q_2=M$ by selecting sufficiently large values of $K_2$ and $H_2$, as in the proof of the optimal replenishment cycle for item 2 in Lemma~\ref{thm:gi:if:reduction}.  We also set $K_1=0$.  (This makes the proof slightly simpler, but is not necessary.  In fact, the reduction works as long as $K_0$ and $K_1$ are chosen such that $K_0 + K_1 = LU$ and $4K_0> (U-L)^2$.)  Thus, the \JRPGI instance becomes:
\begin{align*}
\min\phantom{dd} &\frac{K_0}{\gcd(q,M)} + q \label{eq:gi:redfunc} \\
\text{s.t. \phantom{dd}} & q \in \ZZ^+. \nonumber
 \end{align*}

The convex function $f(q)\equiv K_0/q + q$ (for $q>0$) is a lower bound on the objective function above. By setting $K_0=LU$, this function satisfies $f(L)=f(U)=L+U$ and is minimized at $q=\sqrt{LU}$.

Note that for any integer $q$ satisfying $\gcd(q,M)\neq q$, we have $\gcd(q,M) \leq q/2$. It follows that the objective function value attained by any such value is at least $$\min_{q>0} \frac{2K_0}{q} + q \geq 2\sqrt{2K_0} = 2\sqrt{2LU}> L+U.$$
Now, suppose there exists $q$ in $[L,U]$ dividing $M$. Then, the objective function at such $q$ is at most $\max_{L\leq q\leq U} f(q) = f(L)=L+U$.
On the other hand, if $\JRPMLU \leq L+U$, then the optimizer must be a divisor of $M$. And in this case, $q$ must lie in $[L,U]$.


For the reduction from \NPFact to be sound, we need a representation of $K_0$ of size polynomial in the input of \NPFact (a set of $n$ integers of $B$ bits each). This is not true for $K_0=LU$ as such. However, the reduction is still valid if we use $\ceil{L}$ and $\floor{U}$ instead of $L$ and $U$. Since $U=2^{\lambda(A+0.5)}$ has an exponent no larger than $6Bn$, we need to compute $U$ to $6Bn+1$ bits of precision in order to determine $\floor{U}$. This can be computed in polynomial time~\citep{Brent_Zimmermann_2011}. A similar statement holds for $L$. Thus, $K_0=\ceil{L}\floor{U}$ is of polynomial size.\Halmos
\endproof


This also concludes the proof of Theorem~\ref{thm:gi:hardness}.  The Riemann hypothesis is needed for the result by~\cite{Shor_2011} to apply.

\begin{repeattheorem}\label{thm:gi:hardness}
Assume one can prove the Riemann hypothesis.  Then, the periodic \JRP with two items is NP-hard, under randomized reductions.\\
\end{repeattheorem}

\section{Concluding Remarks}\label{sec:conclusion}
The results in this paper should help to better understand the computational complexity of joint replenishment problems.  We have provided evidence that the simplest multi-item extensions of the \TEOQ model, namely \JRPGI and \JRPGICF, are computationally difficult, even with just two items (in the discrete setting).  Our results suggest that no algorithm may ever be simultaneously exact and polynomial on all inputs; therefore, some kind of concession must be made (e.g., the use of heuristics).

Our technique exploits a natural relation between the problem of coordinating periodic activities and the problem of finding divisors of an integer number. This naturally leads to a notion of hardness based on integer factorization. We not only provide the first integer factorization hardness results, but also show that the \JRPGI is \NP-hard under randomized reductions. However, the assumptions that allow us to obtain this result are more demanding (i.e., Riemann's hypothesis).

It may not be too far-fetched to imagine that similar problems (in Management Science, Operations, and elsewhere), whose computational complexity has remained open, might be classified as integer-factorization hard.  
There is a simple two-step recipe to build these hardness proofs, and they are connected to the only two characteristics needed for this technique to apply: the desire to optimize integer periodic processes, and the incentives these processes have to coordinate.  Both, coordination and periodicity are ubiquitous features in many OM and OR problems.

%
\begin{APPENDIX}{\NPFact Is NP-Complete, Under Randomized Reductions}
For the sake of completeness, we provide a proof of NP-completeness of the \NPFact problem (under randomized reductions).  This proof is based on ideas by \cite{Shor_2011}.   In order to work properly, it requires some number-theoretic assumption.  We use the Riemann hypothesis for this purpose.

The reduction itself is from \textsc{Partition} \citep{Karp_1972}: Given a set of $n$ positive integers, can they be split into two parts so that the sum of the integers in each part is equal?  Note that \textsc{Partition} remains NP-complete even if the integers are restricted to have exactly $B$ bits, with $B=\Omega(\log n)$.  (Here, we use the definition by \cite{HL} of $\Omega$.) Indeed, an instance of the partition problem, $\{a_i\}_{i=1..n}$, in which the largest number has $B$ bits can be transformed into an instance $\{2^{B+\ceil{\log(n)}} + a_i\}_{i=1..n} \cup \{2^{B+\ceil{\log(n)}}\}_{i=1..n}$ with $2n$ integers of $\ceil{\log(n)}+B+1$ bits each. Due to the large coefficient $\approx n2^B$ leading each of the $2n$ terms, equal-sum partitions must have exactly $n$ elements (note that $\sum a_i < n2^B$), and the $a_i$ terms are implicitly split by this partition into two groups with the same sum.

So, consider such an instance of \textsc{Partition}, namely, a set of integers $\{a_i\}_{i=1..n}$, satisfying $a_i \in[2^B, 2^{B+1})$ for some $B=\Omega(\log n)$.    Define $A$ so that $2A = \sum_{i=1}^n a_i$.  Let $\lambda = 3B/2^B$ be a scaling factor so that $b_i \equiv \lambda a_i \in[3B, 6B)$.

Assuming the Riemann hypothesis, \cite{Dudek_Grenie_Molteni_2016} show that for any $x \geq 2$ there are at least $\sqrt{x}$ primes in the interval $(x - 4\sqrt{x} \log x, x+4\sqrt{x} \log x)$. For $x = 2^{b_i}$, this means there are at least $b_i$ prime numbers in $(2^{b_i} - c b_i 2^{b_i/2}, 2^{b_i} + c b_i 2^{b_i/2})$, for some constant $c$, and we can find one such prime, $p_i$, in randomized polynomial time using random sampling.

Note that $cb_i < \lambda 2^{b_i/2}/4n$ for $B=\Omega(\log n)$ sufficiently large. Using the inequalities $2^{-x} \leq 1-x/2$ for $0\leq x \leq 1$ and $e^x \geq 1+x$ for $x\geq -1$, we obtain, in turn, the following inequalities:
\begin{alignat*}{3}
  &2^{b_i} - c b_i 2^{b_i/2} && < p_i && < 2^{b_i} + c b_i 2^{b_i/2},\\
  &2^{b_i} (1- \frac{\lambda}{4n}) &&< p_i && < 2^{b_i}(1 + \frac{\lambda}{4n}),\\
  &2^{b_i} 2^{-\lambda/(2n)} &&< p_i && < 2^{b_i} 2^{\lambda/(2n) }.
\end{alignat*}

Let $L=2^{\lambda (A-0.5) }$ and $U=2^{\lambda (A+0.5)}$. The last inequality implies that if for some subset $S \subseteq \{1, \ldots, n\}$  we have $\sum_{i\in S} a_i = A$, then $L < \Pi_{i\in S} p_i < U$, and vice versa. In other words, $M \equiv \Pi_{i=1}^n p_i$, $L$, $U$ is a yes-instance of \NPFact if and only if $\{a_i\}_{i=1..n}$ is a yes-instance of \textsc{Partition}.  Finally, note that $L$ and $U$ satisfy $2\sqrt{2LU}> L+U$ as long as $(2^{\lambda/2}+2^{-\lambda/2})<2\sqrt{2}$. This is true for $B$ larger than a constant.

\end{APPENDIX}
%
%


\ACKNOWLEDGMENT{The first author gratefully acknowledges the support of the Alexander von Humboldt Foundation and the German Federal Ministry of Education and Research (BMBF). Both authors would like to thank Margarida Carvalho for several useful comments that helped to improve the manuscript.\\}





\bibliographystyle{informs2014}

\end{document}